\documentclass[conference,a4paper]{IEEEtran}

\usepackage[english]{babel}
\usepackage[latin1]{inputenc}
\usepackage{amssymb,amstext, amsfonts}
\usepackage[cmex10]{amsmath}
\usepackage[ruled,vlined]{algorithm2e}
\usepackage{verbatim} 
\usepackage{float}
\usepackage[normalem]{ulem}
\usepackage{color}
\usepackage{cite}

\ifCLASSINFOpdf
   \usepackage[pdftex]{graphicx}
   \DeclareGraphicsExtensions{.pdf,.jpeg,.png}
\else
   \usepackage[dvips]{graphicx}
   \DeclareGraphicsExtensions{.eps}
\fi

\newtheorem{theorem}{Theorem}

\newcommand{\nota}[1]{{\slshape\color{blue}[#1]}}
\renewcommand{\nota}[1]{}
\newcommand{\B}[1]{\boldsymbol{#1}}

\begin{document}

\sloppy


\title{Compressive Classification}



\author{
   \IEEEauthorblockN{Hugo Reboredo, Francesco Renna}
   \IEEEauthorblockA{Instituto de Telecomunica\c{c}\~{o}es\\
    Universidade do Porto, Portugal\\
    Email: \{hugoreboredo,frarenna\}@dcc.fc.up.pt } 
    \and
  \IEEEauthorblockN{Robert Calderbank}
  \IEEEauthorblockA{Department of ECE\\
    Duke University, NC, USA\\
    Email: robert.calderbank@duke.edu} 
   \and
  \IEEEauthorblockN{Miguel R. D. Rodrigues}
  \IEEEauthorblockA{Department of E\&EE\\
    University College London, UK\\
    Email: m.rodrigues@ucl.ac.uk} 
}




\maketitle

\begin{abstract}
This paper derives fundamental limits associated with compressive classification of Gaussian mixture source models. In particular, we offer an asymptotic characterization of the behavior of the (upper bound to the) misclassification probability  associated with the optimal Maximum-A-Posteriori (MAP) classifier that depends on quantities that are dual to the concepts of  diversity gain and coding gain in multi-antenna communications.
The diversity, which is shown to determine the rate at which the probability of misclassification decays in the low noise regime, is shown to depend on the geometry of the source, the geometry of the measurement system and their interplay. The measurement gain, which represents the counterpart of the coding gain, is also shown to depend on geometrical quantities. 
It is argued that the diversity order and the measurement gain also offer an optimization criterion to perform dictionary learning for compressive classification applications.
\end{abstract}

\section{Introduction}

Classification of high dimensional signals is fundamental to the broad fields of signal processing and machine learning. The aim is to increase speed and reliability  while reducing the complexity of discrimination. An approach that has attracted a great deal of current interest is Compressed Sensing (CS) which seeks to capture important attributes of high-dimensional sparse signals from a small set of linear projections. The observation~\cite{Candes06,Donoho06}  that captured the imagination of the signal processing community is that it is possible to guarantee fidelity of reconstruction from random linear projections when the source signal exhibits sparsity with respect to some dictionary. 

Within CS the challenge of signal reconstruction has  attracted the greatest attention, but our focus is different. We are interested in detection rather than estimation, in problems such as hypothesis testing, pattern recognition and anomaly detection that can be viewed as instances of signal classification. It is also natural to employ compressive measurement here since it may be possible to discriminate between signal classes using only partial information about the source signal. The challenge now becomes that of designing measurements that ignore signal features with little discriminative power. In fact, we would argue that the compressive nature of CS makes the paradigm a better fit to classification than to reconstruction.

Compressive classification appears in the machine learning literature as feature extraction or supervised dimensionality reduction. Approaches based on geometrical characterizations of the source have been developed, some like linear discriminant analysis ({LDA}) and principal component analysis ({PCA}) just depending on second order statistics. 
Approaches based on higher-order statistics of the source have also been developed~\cite{Wright09,Liu12,CalderbankJafarpour12,Erdogmus04,Nenadic,CarsonICML12,William12}.
 
In this paper we derive fundamental limits on compressive classification by drawing on measures of operational relevance: the probability of misclassification. We assume that the source signal is described by a Gaussian Mixture Model (GMM) that has been already learned. This assumption is motivated in part by image processing where GMMs have been used very successfully to describe patches extracted from natural images~\cite{Minhua10}. Our main contribution is a characterization of the probability of misclassification as a function of the geometry of the individual classes, their interplay and the number of measurements. We show that the fundamental limits of signal classification are determined by quantities that can be interpreted as the duals of quantities that determine the fundamental limits of multi-antenna communication systems. These quantities include the diversity order and the coding gain which characterize the error probability in multiple input multiple output ({MIMO}) systems in the regime of high signal-to-noise ratio (SNR)~\cite{Tarokh98,Tse05}. We note that wireless communication involves classification rather than reconstruction since the aim is to discriminate the transmitted codeword from the alternative codewords.

We use the following notation: boldface upper-case letters denote matrices (${\bf X}$), boldface lower-case letters denote column vectors (${\bf x}$) and italics denote scalars ($x$); the context defines whether the quantities are deterministic or random. The symbol ${\bf I}$ represents the identity matrix. The operators $\left(\cdot\right)^{T}$, $\mathrm{rank}(\cdot)$,  $\mathrm{det}(\cdot)$ and $\mathrm{pdet}(\cdot)$ represent the 
transpose operator, the rank operator, the determinant operator and the pseudo-determinant operator, respectively. The symbol $\log\left(\cdot\right)$ denotes the natural logarithm. 
For reason of space, we relegate the mathematical proofs of our results to an upcoming journal paper~\cite{Journal}.

\section{The Compressive Classification Problem}
\label{PS}
We consider a classification problem in the presence of compressive and noisy measurements. In particular, we use the standard measurement model given by:
\begin{IEEEeqnarray}{c}
\label{s_model}
{\bf y} =  {\bf \Phi} {\bf x} + {\bf n}
\end{IEEEeqnarray}
where ${\bf y} \in \mathbb{R}^M$ represents the measurement vector, $\mathbf{x} \in \mathbb{R}^N$ represents the source vector, ${\bf \Phi} \in \mathbb{R}^{M \times N}$ represents the measurement matrix and $\mathbf{n} \sim \mathcal{N} \left({\bf 0},\sigma^2 \cdot {\bf I}\right) \in \mathbb{R}^M $ represents standard white Gaussian noise.

We take the measurement matrix to be such that its elements are drawn independently from a zero-mean Gaussian distribution with a certain fixed variance, which is common in various CS problems~\cite{Candes06,Donoho06}. We also take the source signal to follow the well-known GMM, which has been shown to lead to state-of-the-art results in various classification applications including hyper-spectral imaging and digit recognition~\cite{CarsonICML12}. This model assumes that the source signal is drawn from one out of $L$ classes $C_i, i = 1,\ldots,L$, with probability $P_i, i = 1,\ldots,L$, and that the distribution of the source conditioned on $C_i$ is Gaussian with mean $\B{\mu}_i \in \mathbb{R}^N$ and (possibly rank-deficient) covariance matrix ${\bf \Sigma}_i \in  \mathbb{R}^{N \times N}$.

The objective is to produce an estimate of the true signal class given the measurement vector. The Maximum-A-Posteriori (MAP) classifier, which minimizes the probability of misclassification~\cite{duda00}, produces the estimate given by:
\begin{IEEEeqnarray}{LcR}
\hat{C} = {\arg} {\max_{C_i} P\left(C_i \mid {\bf y}\right)} & = &  \arg \max_{C_i} p\left({\bf y} \mid C_i\right) P_i.
\label{class}
\end{IEEEeqnarray}
where $P(C_i | \mathbf{y})$ represents the \textit{a posteriori} probability of class $C_i$ given the measurement vector $\mathbf{y}$  and $p(\mathbf{y} | C_i)$ represents the probability density function of the measurement vector $\mathbf{y}$ given the class $C_i$.


We base the analysis -- in line with the standard practice in multiple-antenna communications systems~\cite{Tarokh98,Tse05}  --  on an upper bound to the probability of misclassification of the MAP classifier $\bar{P}_{err}$ , rather than the exact probability of misclassification $P_{err}$. We also base the analysis on two fundamental metrics that characterize the asymptotic performance of the upper bound to the probability of misclassification in the low noise regime, which is relevant to various emerging classification tasks~\cite{CarsonICML12}. In particular, we define the diversity order of the measurement model in~(\ref{s_model}) as:
\begin{equation}
	d = \lim_{\sigma^2 \to 0} \frac{\log\bar{P}_{err}(\sigma^2)}{\log{\sigma^2}},
\label{diversity}	
\end{equation}
that determines how (the upper bound to) the misclassification probability decays at low noise levels~\cite{ZheTse03,Paul03}. We also define the measurement gain of the measurement model in~(\ref{s_model}) as:
\begin{IEEEeqnarray}{LcR}
{g_m} =  \lim_{\sigma^2 \rightarrow 0}  \sigma^2 \cdot \frac{1}{ \sqrt[d]{\bar{P}_{err}(\sigma^2)}  } ,
\label{gm}
\end{IEEEeqnarray}
that determines the offset of (the upper bound to) the misclassification error probability at low noise levels. These quantities admit a counterpart in multiple-antenna communications -- for example, the measurement gain corresponds to the standard coding gain. It turns out that the behavior of the upper bound to the misclassification probability mimics closely the behavior of the exact misclassification probability -- as shown in the sequel -- bearing witness to the value of the approach.

The characterization of the performance measures in (\ref{diversity}) and (\ref{gm}) will be expressed via quantities that relate to the geometry of the measurement model, namely, the rank and the pseudo-determinant of certain matrices. In particular, we define the behavior of (\ref{diversity}) and (\ref{gm}) via the geometry of the linear transformation of the source signal effected by the measurement ``channel'', by using the quantities:

\begin{itemize}

\item $r_i = \mathrm{rank} (\mathbf{\Phi} \mathbf{\Sigma}_i \mathbf{\Phi}^T)$ and $v_i = \mathrm{pdet} (\mathbf{\Phi} \mathbf{\Sigma}_i \mathbf{\Phi}^T)$, which measure the dimension and volume, respectively, of the sub-space spanned by the linear transformation of the signals in class $C_i$;

\item $r_{ij} = \mathrm{rank} (\mathbf{\Phi} (\mathbf{\Sigma}_i + \mathbf{\Sigma}_j) \mathbf{\Phi}^T)$ and $v_{ij} = \mathrm{pdet} (\mathbf{\Phi} (\mathbf{\Sigma}_i + \mathbf{\Sigma}_j) \mathbf{\Phi}^T)$, which measure the dimension and volume, respectively, of the union of sub-spaces spanned by the linear transformation of the signals in classes $C_i$ or $C_j$.
\end{itemize}

We also define the behavior of (\ref{diversity}) and (\ref{gm}) via the geometry of the original source signal, by using the quantities:

\begin{itemize}

\item $r_{\mathbf{\Sigma}_i} = \mathrm{rank}(\mathbf{\Sigma}_i)$, which relates to the dimension of the sub-space spanned by input signals in $C_i$;

\item $r_{\mathbf{\Sigma}_{ij}} = \mathrm{rank}(\mathbf{\Sigma}_i +  \mathbf{\Sigma}_j)$, which relates to the dimension of the union of sub-spaces spanned by input signals in $C_i$ or $C_j$.

\end{itemize}

We argue that this two-step approach casts further insight into the characteristics of the compressive classification problem, by allowing us to untangle in a systematic manner the effect of the measurement matrix and the effect of the source geometry.


\section{The Case of Two Classes}
\label{two_class}

We now consider a two-class compressive classification problem. The Bhattacharyya bound, which represents a specialization of the Chernoff bound~\cite{Bhat43}, leads to an upper bound to the probability of misclassification given by~\cite{duda00}:
\begin{equation}
\label{P_err_Bhat}
\bar{P}_{err} = \sqrt{P_1P_2}~ e^{-K\left(1,2\right)},
\end{equation}
where
\begin{IEEEeqnarray}{lcl}
\label{exp_Bhat}
{K\left(i,j\right)}  =    \nonumber \\
 \frac{1}{8}{\textstyle\left[{\bf \Phi}\left(\boldsymbol{\mu}_i - \boldsymbol{\mu}_j\right)\right]^T \left[\frac{{\bf \Phi}\left({\bf \Sigma}_i+{\bf \Sigma}_j\right){\bf \Phi}^T + \sigma^2 \mathbf{I}}{2}\right]^{-1} \left[{\bf \Phi}\left(\boldsymbol{\mu}_i - \boldsymbol{\mu}_j\right)\right]  } \nonumber \\ 
{\displaystyle + \frac{1}{2}\log\frac{\mathrm{det} \left(\frac{{\bf \Phi}\left({\bf \Sigma}_i+{\bf \Sigma}_j\right){\bf \Phi}^T + \sigma^2 \mathbf{I}}{2}\right)}{\sqrt{\mathrm{det} \left({\bf \Phi}{\bf \Sigma}_i{\bf \Phi}^T + \sigma^2 \mathbf{I}\right)  \mathrm{det} \left({\bf \Phi}{\bf \Sigma}_j{\bf \Phi}^T + \sigma^2 \mathbf{I}\right)}} }.
\end{IEEEeqnarray}
The Bhattacharyya based upper bound to the probability of misclassification encapsulated in (\ref{P_err_Bhat}) and (\ref{exp_Bhat}) is the basis of the ensuing analysis. This analysis treats the case where the classes are zero-mean, i.e. $\B{\mu}_1 = \B{\mu}_2 = \B{0}$, and the case where classes are non-zero mean, i.e. $\B{\mu}_1 \neq \B{0}$ or $\B{\mu}_2 \neq \B{0}$, separately. The zero-mean case exhibits the main operational features of the compressive classification problem; the nonzero-mean case exhibits occasionally additional operational features, e.g. infinite diversity order.

\subsection{Zero-Mean Classes}

The following Theorem offers a view of the asymptotic behavior of the probability of misclassification for the two-class compressive classification problem with zero-mean classes, by leveraging directly the geometry of the linear transformation of the source signal effected by the measurement ``channel''.

\vspace{0.30cm}
\begin{theorem}
\label{theorem1}
Consider the measurement model in \eqref{s_model} where ${\bf x} \sim \mathcal{N} ({\bf 0},{\bf \Sigma}_1)$ with probability $P_1$ and ${\bf x} \sim \mathcal{N} ({\bf 0},{\bf \Sigma}_2)$ with probability $P_2 = 1 - P_1$. Then, the upper bound to the probability of misclassification in \eqref{P_err_Bhat} behaves as:
\begin{itemize}
  \item	If $\frac{r_1+r_2}{2} = r_{12}$ then,
\end{itemize}
				\begin{equation}
				\bar{P}_{err} =  \mathcal{O}\left(1\right),~~\sigma^2 \to 0
				\end{equation}
\begin{itemize}
  \item If $\frac{r_1+r_2}{2} < r_{12}$ then,
\end{itemize}
				\begin{equation}
				\bar{P}_{err} = \left(\frac{g_m}{\sigma^2} \right) ^ {- d} + o \left(\left(\frac{1}{\sigma^2}\right)^{- d}\right),~~\sigma^2 \to 0
				\end{equation}
where
\begin{equation}
\label{diver}
	d = -\frac{1}{2}\left(\frac{r_1+r_2}{2}- r_{12}\right)
\end{equation}
and 
\begin{equation}
\label{measure}
{g_m} = 
{ \left[\sqrt{P_1P_2}\left[\frac{v_{12}}{\sqrt{v_1 v_2}}\right]^{-\frac{1}{2}}\right]^{-\frac{1}{d}}.
}
\end{equation}
\end{theorem}

The following Theorem now describes the asymptotic behavior of the probability of misclassification for the two-class compressive classification problem with zero-mean classes, by leveraging instead the geometry of the source signals. 
The result uses the fact that $N \geq r_{{\bf \Sigma}_{12}} \geq \max\left(r_{{\bf \Sigma}_1},r_{{\bf \Sigma}_2}\right)$ and, with  probability 1, $r_1 = \min\left(M,r_{{\bf \Sigma}_1}\right)$, $r_2= \min\left(M,r_{{\bf \Sigma}_2}\right)$ and 
$r_{12} = \min\left(M,r_{{\bf \Sigma}_{12}}\right)$. The result also assumes, without loss of generality, that $r_{{\bf \Sigma}_1} \leq r_{{\bf \Sigma}_2}$.

\vspace{0.30cm}
\begin{theorem}
\label{theorem2}
Consider the measurement model in \eqref{s_model} where ${\bf x} \sim \mathcal{N} ({\bf 0},{\bf \Sigma}_1)$ with probability $P_1$ and ${\bf x} \sim \mathcal{N} ({\bf 0},{\bf \Sigma}_2)$ with probability $P_2 = 1 - P_1$. Then, the upper bound to the probability of misclassification in \eqref{P_err_Bhat} behaves as:
\begin{itemize}
  \item If $M \leq r_{{\bf \Sigma}_1} \leq r_{{\bf \Sigma}_2} \leq r_{{\bf \Sigma}_{12}}$, or $r_{{\bf \Sigma}_1} \leq r_{{\bf \Sigma}_2} \leq r_{{\bf \Sigma}_{12}} \leq M$ and $\frac{r_{{\bf \Sigma}_1}+r_{{\bf \Sigma}_2}}{2} = r_{{\bf \Sigma}_{12}}$ then, 
\end{itemize}
								\begin{equation}
								\bar{P}_{err} =  \mathcal{O}\left(1\right),~~\sigma^2 \to 0
								\end{equation}
\begin{itemize}
  \item Otherwise, 
\end{itemize}
								\begin{equation}
								\bar{P}_{err} = \left(\frac{g_m}{\sigma^2} \right) ^ {- d} + o \left(\left(\frac{1}{\sigma^2}\right)^{- d}\right),~~\sigma^2 \to 0
								\end{equation}
								where
\begin{equation}
{g_m} = 
{ \left[\sqrt{P_1P_2}\left[\frac{v_{12}}{\sqrt{v_1 v_2}}\right]^{-\frac{1}{2}}\right]^{-\frac{1}{d}}
}
\label{gm2}
\end{equation}
and when $r_{{\bf \Sigma}_1} < M \leq r_{{\bf \Sigma}_2} \leq r_{{\bf \Sigma}_{12}}$:
\begin{equation}
d = -\frac{1}{2}\left(\frac{r_{{\bf \Sigma}_1} - M}{2}\right);
\end{equation}
when $r_{{\bf \Sigma}_1} \leq r_{{\bf \Sigma}_2} < M < r_{{\bf \Sigma}_{12}}$:
\begin{equation}
d = -\frac{1}{2}\left(\frac{r_{{\bf \Sigma}_1}+r_{{\bf \Sigma}_2}}{2}- M\right);
\end{equation}
and when $r_{{\bf \Sigma}_1} \leq r_{{\bf \Sigma}_2} \leq r_{{\bf \Sigma}_{12}} \leq M$ and $\frac{r_{{\bf \Sigma}_1}+r_{{\bf \Sigma}_2}}{2} < r_{{\bf \Sigma}_{12}}$:
\begin{equation}
\label{max_diver}
d = -\frac{1}{2}\left(\frac{r_{{\bf \Sigma}_1}+r_{{\bf \Sigma}_2}}{2}- r_{{\bf \Sigma}_{12}}\right).
\end{equation}
\end{theorem}

It is now instructive to probe further on the characterizations embodied in Theorems \ref{theorem1} and \ref{theorem2} to infer the main operational features of the two-class compressive classification problem.

The characterization encapsulated in Theorem \ref{theorem1} admits a very simple interpretation:

\begin{itemize}

\item if $\frac{\left(r_1 + r_2\right)}{2} = r_{12}$, then the sub-spaces spanned by the signals in classes 1 and 2 overlap completely -- the upper bound to the misclassification probability exhibits an error floor because it is not possible to distinguish the classes perfectly as the noise level approaches zero;

\item if $\frac{\left(r_1 + r_2\right)}{2} < r_{12}$ then the sub-spaces spanned by the signals in classes 1 and 2 do not overlap completely -- the upper bound to the misclassification error probability (and the true error probability) then does not exhibit an error floor as it is possible to distinguish the classes perfectly as the noise level approaches zero. The lower the degree of overlap, the higher the diversity order -- this is measured via the interplay of the various ranks, $r_1$, $r_2$ and $r_{12}$;

\item the scenario $\frac{\left(r_1 + r_2\right)}{2} > r_{12}$ is not possible in view of the geometry of the two-class problem.

\end{itemize}

On the other hand, the characterization encapsulated in Theorem \ref{theorem2} offers the means to articulate about the interplay between the number of measurements and the source geometry. Of particular importance:
\begin{itemize}

\item if $M \leq r_{{\bf \Sigma}_{1}}$, the upper bound will exhibit an error floor a low noise levels; conversely, if $M > r_{{\bf \Sigma}_{1}}$ and $\frac{r_{{\bf \Sigma}_{1}} + r_{{\bf \Sigma}_{2}}} {2} < r_{{\bf \Sigma}_{12}}$ the upper bound will not exhibit such an error floor at low noise levels;

\item in addition, if $M \geq r_{{\bf \Sigma}_{12}}$ additional measurements will have no impact on diversity order.

\end{itemize}

Overall, it is possible to argue that the diversity order is a function of the difference between the sub-spaces associated with the two classes, which is given by \eqref{diver}: by gradually increasing the number of measurements from 1 up to $r_{{\bf \Sigma}_{12}}$ it is possible to extract the highest diversity level equal to \eqref{max_diver}; however, increasing the number of measurements past $r_{{\bf \Sigma}_{12}}$ does not offer a higher diversity level -- instead, it only translates into a higher measurement gain. One then understands the role of measurement as a way to probe the differences between the classes.

In contrast, the measurement gain is a function of the exact geometry of the classes in the Gaussian mixture model. It increases with the ratio of the product of the non-zero eigenvalues of  ${\bf \Phi}\left({\bf \Sigma}_1+{\bf \Sigma}_2\right){\bf \Phi}^T$ to the product of the singular values of ${\bf \Phi}{\bf \Sigma}_1{\bf \Phi}^T$ and ${\bf \Phi}{\bf \Sigma}_2{\bf \Phi}^T$. 

We note that there is often flexibility in the definition of the properties of signal classes of a GMM, i.e. the dictionary~\cite{Minhua10}. Measurement gain, and to a certain extent the diversity gain, can then provide an optimization criterion for dictionary\nota{check} design for compressive classification applications. 

\subsection{Nonzero-mean classes}

The following Theorem generalizes the description of the asymptotic behavior of the probability of misclassification from the zero-mean to the nonzero-mean two-class compressive classification problem.

\vspace{0.30cm}

\begin{theorem}
\label{theo:nonzero}
Consider the measurement model in (\ref{s_model}) where $\mathbf{x} \sim \mathcal{N} (\B{\mu}_1,\mathbf{\Sigma}_1)$ with probability $P_1$ and $\mathbf{x} \sim \mathcal{N} (\B{\mu}_2,\mathbf{\Sigma}_2)$ with probability $P_2 = 1 - P_1$ and $\B{\mu}_1 \neq \B{\mu}_2 \neq \B{0}$. If
\begin{equation}
  \mathrm{im}(\B{\Phi} (\B{\mu}_1 - \B{\mu}_2)  (\B{\mu}_1 - \B{\mu}_2)^{T} \B{\Phi}^{T} ) \not \subseteq \mathrm{im}( \mathbf{\Phi} (\mathbf{\Sigma}_1 + \mathbf{\Sigma}_2) \mathbf{\Phi}^{T}),
  \label{eq:im}
\end{equation}
then the upper bound to the probability of misclassification in (\ref{P_err_Bhat}) decays exponentially with $1/\sigma^2$ as $\sigma^2 \rightarrow 0$; otherwise,
\begin{equation}
								\bar{P}_{err} = \left(\frac{a \cdot g_m}{\sigma^2} \right) ^ {- d} + o \left(\left(\frac{1}{\sigma^2}\right)^{- d}\right),~~\sigma^2 \to 0
								\end{equation}
where $a > 1 $ is a finite constant which depends on the first term in~(\ref{exp_Bhat}), and $g_m$ and $d$ are as in Theorems \ref{theorem1} and \ref{theorem2}.
\end{theorem}

The characterization embodied in Theorem~{\ref{theo:nonzero}} illustrates that the asymptotic behavior of the upper bound of the error probability for classes with non-zero mean can be different from that for classes with zero mean. The differences in behavior trace back to the fact that $M >r_{\mathbf{\Sigma}_{12}}$ represents a necessary condition for condition (\ref{eq:im}) to hold. In the non-zero mean case, choosing $M >r_{\mathbf{\Sigma}_{12}}$ leads to a diversity order $d = \infty$; in contrast,  in the zero-mean case choosing $M >r_{\mathbf{\Sigma}_{12}}$ does not affect the diversity order. Letting $M  \leq r_{\mathbf{\Sigma}_{12}}$ induces the same diversity order both for nonzero-mean and zero-mean classes. The presence of the nonzero-mean here then impacts only the measurement gain.

One concludes that in the non-zero mean case, increasing the number of measurements past $r_{\mathbf{\Sigma}_{12}}$ can have a dramatic effect on the performance. 
Geometrically, this result reflects the fact that, when embedded in a higher dimensional space ($\mathbb{R}^M$ in our cases), the affine spaces corresponding to the classes are  separated when $\sigma^2 \rightarrow 0$. 

\section {The Case of Multiple Classes}
\label{mul_class}

We now consider a multiple class compressive classification problem, where $L \geq 3$. The generalization of the two-class results to the multiple-class case is possible by using the union bound in conjunction with the two-class Bhattacharyya bound.

The combination of the union bound with Bhattacharyya bound leads immediately to an upper bound to the probability of misclassification given by ~\cite{Wim12}:
\begin{IEEEeqnarray}{LcR}
\bar{P}_{err} = \sum_{i=1}^L \sum_{\substack{
   j = 1 \\
   j \neq i
  }}^L \sqrt{P_iP_j} e^{-K\left(i,j\right)} P_i
  \label{multiclass}
\end{IEEEeqnarray}
and $K(i,j)$ is given by \eqref{exp_Bhat}.

The fact that the form of the upper bound in (\ref{multiclass}) is akin to the form of the upper bound in (\ref{P_err_Bhat}), involving only in addition various pair-wise misclassification terms that capture the interaction between the different classes, leads to the immediate generalization of the results encapsulated in the previous Theorems.

In particular, we can argue that the upper bound to the misclassification probability will exhibit an error floor if at least one of the pair-wise misclassification probabilities also exhibits an error floor. Conversely, the misclassification probability will tend to zero as $\sigma^2$ tends to zero if all the pairwise misclassification probabilities also tend to zero.

The diversity order of the multiple-class misclassification probability now corresponds to the lowest diversity order of the pairwise 
misclassification probabilities. Similarly, the measurement gain of the multiple-class misclassification probability corresponds to the measurement gain of the pairwise misclassification probability associated associated with the lowest diversity order. Therefore, it is possible to capitalize on the results embodied in Theorems \ref{theorem1}, \ref{theorem2} and \ref{theo:nonzero} to understand the behavior of the multiple-class compressive classification problem immediatly.

\section{Numerical Results}
\label{num_res}

\begin{figure}[tb]
   \centering
   \includegraphics[width=3.5in, height=2in]{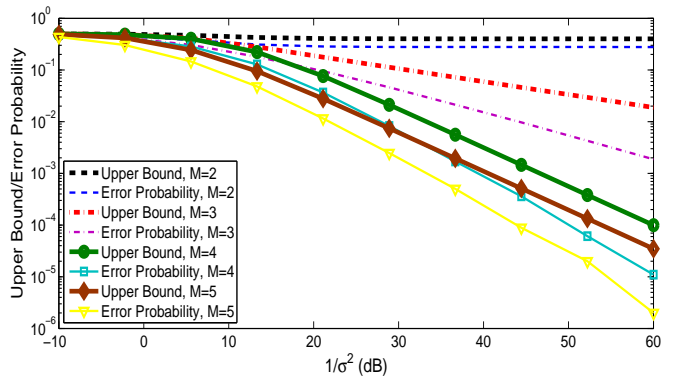}
\vspace{-0.65cm}
   \caption{Upper bound to error probability and true error probability \emph{vs.} $1/\sigma^2$ for different number of measurements $M$, for the zero-mean two-class compressive classification problem. ($r_{{\bf \Sigma}_1} = 2$,  $r_{{\bf \Sigma}_2} = 3$,  $r_{{\bf \Sigma}_{12}}=4$, $\boldsymbol{\mu}_1 = \boldsymbol{\mu}_2 = {\bf 0}$ and $N=6$).}
   \label{fig:two_class_1}
\vspace{-0.30cm}
 \end{figure}

\begin{figure}[tb]
   \centering
   \includegraphics[width=3.5in, height=2in]{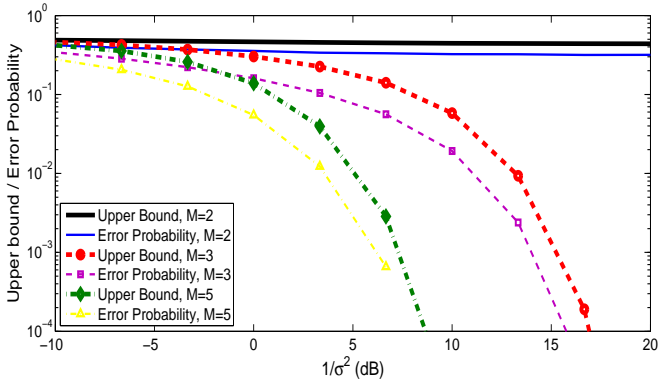}
   \vspace{-0.65cm}
   \caption{Upper bound to error probability and true error probability \emph{vs.} $1/\sigma^2$ for different number of measurements $M$, for a non-zero-mean two-class compressive classification problem. ($r_{{\bf \Sigma}_1} = 2$,  $r_{{\bf \Sigma}_2} = 2$,  $r_{{\bf \Sigma}_{12}}=2$, $\boldsymbol{\mu}_1 \neq \boldsymbol{\mu}_2 \neq {\bf 0}$ and $N=6$).}
   \label{fig:mu_nonzero}
   \vspace{-0.30cm}
 \end{figure}

\begin{figure}[t]
   \centering
   \includegraphics[width=3.5in, height=2in]{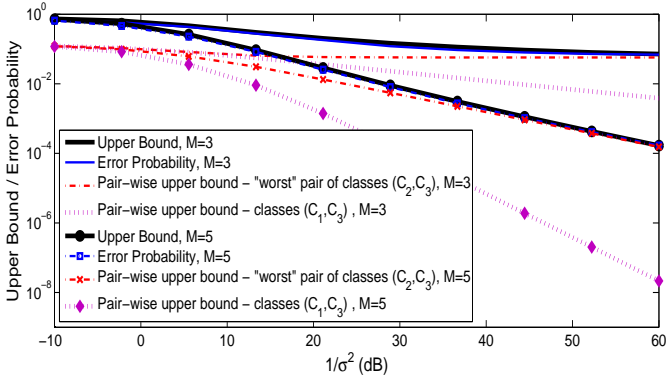}
   \vspace{-0.65cm}
   \caption{Upper bound to error probability and true error probability \emph{vs.} $1/\sigma^2$ for different number of measurements $M$, for a multiple class classification problem. ($r_{{\bf \Sigma}_1} = 2$, $r_{{\bf \Sigma}_2} = 3$, $r_{{\bf \Sigma}_3} = 3$, $r_{{\bf \Sigma}_4} = 2$, $r_{{\bf \Sigma}_{12}} = 4$, $r_{{\bf \Sigma}_{13}} = 5$, $r_{{\bf \Sigma}_{14}} = 4$, $r_{{\bf \Sigma}_{23}} = 4$, $r_{{\bf \Sigma}_{24}} = 5$, $r_{{\bf \Sigma}_{34}} = 4$, $\boldsymbol{\mu}_i = \B{0}, i=1,\cdots,4$ and $N=6$).}
   \label{fig:mul_class}
   \vspace{-0.30cm}
 \end{figure}

We now present a series of results that illustrate the main operational features of compressive classification of Gaussian mixture models. In particular, we consider three experiments where the exact covariance matrices and/or mean vectors of the classes have been generated randomly:

\begin{itemize}

\item A two-class compressive classification problem with zero-mean classes where $L = 2$, $r_{{\bf \Sigma}_1} = 2$,  $r_{{\bf \Sigma}_2} = 3$,  $r_{{\bf \Sigma}_{12}}=4$, $\B{\mu}_1 = \B{\mu}_2 = \B{0}$ and $N=6$ (see Figure \ref{fig:two_class_1});

\item A two-class compressive classification problem with nonzero-mean classes where $L = 2$, $r_{{\bf \Sigma}_1} = 2$,  $r_{{\bf \Sigma}_2} = 2$,  $r_{{\bf \Sigma}_{12}}=2$, $\boldsymbol{\mu}_1 \neq \boldsymbol{\mu}_2 \neq {\bf 0}$ and $N=6$ (see Figure \ref{fig:mu_nonzero});

\item A multi-class classification problem where $L= 4$, $r_{{\bf \Sigma}_1} = 2$, $r_{{\bf \Sigma}_2} = 3$, $r_{{\bf \Sigma}_3} = 3$, $r_{{\bf \Sigma}_4} = 2$, $r_{{\bf \Sigma}_{12}} = 4$, $r_{{\bf \Sigma}_{13}} = 5$, $r_{{\bf \Sigma}_{14}} = 4$, $r_{{\bf \Sigma}_{23}} = 4$, $r_{{\bf \Sigma}_{24}} = 5$, $r_{{\bf \Sigma}_{34}} = 4$, $\boldsymbol{\mu}_i = {\bf 0}, i=1,\cdots,L$ and $N=6$ (see Figure \ref{fig:mul_class}).

\end{itemize}

The results confirm the asymptotic behavior of the upper bound to the misclassification probability (as well as the true misclassification probability) unveiled in Theorems \ref{theorem1} -- \ref{theo:nonzero}.

Figure \ref{fig:two_class_1} shows that, for $M=1,2$, the upper bound exhibits an error floor, as expected. As $M$ increases ($M=3,4$) we can observe that i) it becomes possible to drive the upper bound to the misclassification probability to zero at low noise levels, and therefore perform classification without errors; and ii) the diversity gain also increases with $M$. We can also verify that if $M \geq r_{{\bf \Sigma}_{12}}$, having more measurements does not increase the diversity gain, but yields a higher measurement gain. 

Figure \ref{fig:mu_nonzero} shows that, when $M \leq r_{{\bf \Sigma}_{12}}$, the upper bound exhibits an error floor, as expected. When $M > r_{{\bf \Sigma}_{12}}$, the upper bound will decay exponentially to zero (at low noise levels) as presented in Theorem~\ref{theo:nonzero}.

In Figure \ref{fig:mul_class} we can observe that the upper bound to the misclassification probability is, indeed, dominated by the behavior of the worst pair of classes (which in such scenario corresponds to the pair ($C_2,C_3$), yielding a maximum diversity of $d=1$, for $M \geq 4$), by comparing with any other pair-wise upper bound, (e.g. the behavior of the pair ($C_1,C_3$) also depicted in the figure).

\section{Conclusion}
\label{conclusions} 
 
This paper studies fundamental limits in compressive classification of Gaussian mixture models. In particular, it is shown that the asymptotic behavior of (the upper bound to) the misclassification probability, which is intimately linked to the geometrical properties of the source and the measurement system, also captures well the behavior of the true misclassification probability. Moreover, it is recognized that the key quantities that determine the asymptotic behavior of the misclassification probability are akin to standard quantities used to characterize the behavior of the error probability in multiple-antenna communications: diversity and coding gain.

The practical relevance of the results -- and beyond the theoretical insight -- relates to the possibility of integrating the asymptotic characterizations with dictionary learning methods for compressive classification. The diversity order and the measurement gain -- in view of its links to the geometry of the measurement system and the geometry of the source -- offer a means to pose optimization problem that offer an opportunity to construct dictionaries with good discriminative power.

\bibliographystyle{IEEEtran}
\bibliography{IEEEabrv,cl}

\end{document}